\documentclass[aps,nofootinbib,preprint]{revtex4}

\usepackage{graphicx}
\usepackage{graphics}
\usepackage{amsmath}
\usepackage{subfigure}

\begin{document}
\title{Constraints on the time variation of the fine structure constant by the 5-year WMAP data}

\author{Masahiro Nakashima $^{1,2}$}
\author{Ryo Nagata $^{2}$}
\author{Jun'ichi Yokoyama $^{2,3}$}

\affiliation{$^{1}$ Department of Physics, Graduate School of Science,\\  
The University of Tokyo, Tokyo 113-0033, Japan \\
$^{2}$ Research Center for the Early Universe (RESCEU), \\
Graduate School of Science, The University of Tokyo, Tokyo 113-0033, Japan \\
$^{3}$ Institute for the Physics and Mathematics of the Universe(IPMU), \\
The University of Tokyo, Kashiwa, Chiba, 277-8568, Japan}

\begin{abstract}
The constraints on the time variation of the fine structure constant at recombination epoch relative to its present value, $\Delta\alpha/\alpha \equiv (\alpha_{\mathrm{rec}} - \alpha_{\mathrm{now}})/\alpha_{\mathrm{now}}$, are obtained from the analysis of the 5-year WMAP cosmic microwave background data. As a result of Markov-Chain Monte-Carlo analysis, it is found that, contrary to the analysis based on the previous WMAP data, the mean value of $\Delta\alpha/\alpha=-0.0009$ does not change significantly whether we use the Hubble Space Telescope (HST) measurement of the Hubble parameter as a prior or not. The resultant 95$\%$ confidence ranges of $\Delta\alpha/\alpha$ are $-0.028 < \Delta\alpha/\alpha < 0.026$ with HST prior and $-0.050 < \Delta\alpha/\alpha < 0.042$ without HST prior.
\end{abstract}

\begin{flushright}
RESCEU-61/08
\end{flushright}

\maketitle

 The variation of the fundamental physical constants is a longstanding issue. Dirac first considered such a possibility \cite{Dirac:1937ti,Dirac:1938mt} and proposeded that the Newton constant should be inversely proportionaol to time. While his claim is not compatible with the current observations, recent unification theories such as superstring theories naturally predict the variation of the fundamental constants \cite{Uzan:2002vq}. Because of these theoretical motivations, it is important to measure their possible time variation observationally. 

 Among various fundamental constants, the time variation of the fine structure constant $\alpha$ has been most extensively discussed in observational contexts. We briefly summarize those terrestrial and celestial limits on $\alpha$ as follows \cite{Chiba:2001ui}. 

\begin{itemize}
\item The atomic clocks constrain the current value of the temporal derivative of the fine structure constant as $\dot{\alpha}/\alpha = (-3.3\pm 3.0)\times 10^{-16} \mathrm{yr}^{-1}$ \cite{Blatt:2008su,Fortier:2007jf,Peik:2004qn,Fischer:2004jt}.
\item $\dot{\alpha}/\alpha=(-1.6\pm 2.3) \times 10^{-17} \mathrm{yr}^{-1}$ from the measurement of the frequency ratio of aluminium and mercury single-ion optical clocks \cite{Rosenband:2008}. 
\item $\Delta\alpha/\alpha = (-0.8\pm 1.0) \times 10^{-8}$ or $\Delta\alpha/\alpha=(0.88 \pm 0.07) \times 10^{-7}$ from the Oklo natural reactor in Gabon ($z \sim 0.1$) \cite{Fujii:2002hc}. 
\item $\Delta\alpha/\alpha = (-0.57\pm 0.11) \times 10^{-5} \ (z \sim 0.2 $-$4.2$) \cite{Murphy:2004}, and $\Delta\alpha/\alpha = (-0.64 \pm 0.36)\times 10^{-5} \ (z\sim 0.4$-$2.3)$ \cite{Murphy:2006vs,Murphy:2007qs} from spectra of quasars, the former  of which is from the Keck/HIRES instrument, and the latter from the Ultraviolet and Visual Echelle Spectrograph (UVES) instrument.
\item $-5.0\times 10^{-2} < \Delta\alpha/\alpha < 1.0 \times 10^{-2} \ (95 \% \mathrm{C.L.})$ from
big bang nucleosynthesis (BBN, $z\sim 10^{9}$-$10^{10}$) \cite{Ichikawa:2002bt}.
\item $-0.048 <\Delta\alpha/\alpha < 0.032$ \cite{Ichikawa:2006nm}, $-0.06 < \Delta\alpha/\alpha < 0.01$ \cite{Rocha:2003gc} or $-0.039 < \Delta\alpha/\alpha < 0.010$ \cite{Stefanescu:2007aa} ($95\% \mathrm{C.L.}$) from the cosmic microwave background (CMB, $z\sim 10^{3}$), the former two of which are based on the analysis of the 1-year WMAP data and the last one on the 3-year WMAP data. 
\end{itemize}
We also note that the seasonal variation effect on $\alpha$ has been also discussed in \cite{Barrow:2008se}.

 In this paper, we focus on the CMB constraint on $\alpha$ from 5-year WMAP data, finding new limits on its value at the recombination epoch. While the other physical constants may vary in time simultaneously, they are so model-dependent \cite{Chiba:2006xx} that we consider only the variation of $\alpha$ here. An example of such a class of models can be found in \cite{Bekenstein:1982eu,Sandvik:2001rv}.

 Both CMB and BBN \cite{Olive:1999ij} are useful in order to obtain the constraints of the variation of $\alpha$ over a cosmological time scale. Although BBN is surperior in that it can probe a longer timespan, it has a drawback that the effect of $\alpha$ in Helium abundance $Y_{p}$ is model dependent so that we cannot get a robust result from BBN analysis. On the other hand, the physics of the CMB is much simpler and well understood with high precision data, so we can obtain a meaningful limit on the variation of $\alpha$ from the CMB data.
 
 As is well known, changing the value of the fine structure constant affects the CMB power spectrum mainly through the change of the epoch of recombination \cite{Hannestad:1998xp,Kaplinghat:1998ry}. Hence it probes the value of $\alpha$ at this particular epoch. Let us summerize the main part of the recombination process and see how $\alpha$ appears.

 Following the treatments of \cite{Seager:1999bc, Wong:2007ym}, which are implemented in the RECFAST code \cite{scott}, recombination process is approximated by the evolutions of three variables: the proton fraction $x_{p}=n_{p}/n_{\mathrm{H}}$, the fraction of the singly ionized Helium $x_{\mathrm{HeII}}=n_{\mathrm{HeII}}/n_{\mathrm{H}}$, and the matter temperature $T_{M}$. Here, $n_{\mathrm{H}}$ is defined as the total Hydrogen number density. Their evolution equations read
\begin{align}
\frac{dx_{p}}{dz} & =\frac{C_{\mathrm{H}}}{H(z)(1+z)} \left[ \alpha_{\mathrm{H}} x_{e}x_{p}n_{\mathrm{H}} - \beta_{\mathrm{H}}(1-x_{e})\exp\left( -\frac{h\nu_{\mathrm{\mathrm{H}}}}{k_{B}T_{M}}\right) \right], \\
  \frac{dx_{\mathrm{HeII}}}{dz} & =\frac{C_{\mathrm{He}}}{H(z)(1+z)} \left[\alpha_{\mathrm{HeI}} x_{\mathrm{HeII}}x_{e}n_{\mathrm{H}} - \beta_{\mathrm{He}}(f_{\mathrm{He}}-x_{\mathrm{HeII}})\exp\left( -\frac{h\nu_{\mathrm{He2^{1}s}}}{k_{B}T_{M}}\right) \right], \\
  \frac{dT_{M}}{dz} &= \frac{8\sigma_{T}a_{R}T_{R}^{4}}{3H(z)(1+z)m_{e}}\frac{x_{e}}{1+f_{\mathrm{He}}+x_{e}}\left( T_{M}-T_{R}\right)+\frac{2T_{M}}{1+z}. 
\end{align}  
In the above equations, $H(z)$ is the Hubble expansion rate at the redshift $z$, $\sigma_{T} = 2\alpha^{2}h^{2}/(3\pi m_{e}^{2} c^{2})$ is the Thomson scattering cross section, $a_{R}= k_{B}^{4}/(120\pi c^{3} h^{3})$ is the blackbody constant\footnote{$h$ is the Planck constant, $k_{B}$ is the Boltzmann constant, $c$ is the speed of light.}, $\nu_{\mathrm{H}}=c/(121.5682\ \mathrm{nm})$ is the Ly-$\alpha$ frequency, $\nu_{\mathrm{He2^{1}s}}=c/(60.1404\ \mathrm{nm})$ is the He$2^{1}s$-$1^{1}s$ frequency, $x_{e}=n_{e}/n_{\mathrm{H}}=x_{p} + x_{\mathrm{HeII}}$ is the free electron fraction, $T_{R}$ is the radiation temperature, and $f_{\mathrm{He}}=Y_{p}/(4(1-Y_{p}))$ is the number ratio of Helium to Hydrogen, where $Y_{p}$ is the primordial Helium mass fraction which we take 0.24. $C_{\mathrm{H}}$ ($C_{\mathrm{He}}$) is the so-called Peebles reduction factors 
\begin{align}
C_{\mathrm{H}} &= \frac{1+ K_{\mathrm{H}}\Lambda_{\mathrm{H}}n_{\mathrm{H}}(1-x_{p})}{1+K_{\mathrm{H}}(\Lambda_{\mathrm{H}}+\beta_{\mathrm{H}})n_{\mathrm{H}}(1-x_{p})}, \\
C_{\mathrm{He}} &= \frac{1+ K_{\mathrm{He}}\Lambda_{\mathrm{He}}n_{\mathrm{H}}(f_{\mathrm{He}}-x_{\mathrm{HeII}})\exp\left(-h\nu_{ps}/k_{B}T_{M}\right)}{1+K_{\mathrm{He}}(\Lambda_{\mathrm{He}}+\beta_{\mathrm{He}})n_{\mathrm{H}}(f_{\mathrm{He}}-x_{\mathrm{HeII}})\exp\left(-h\nu_{ps}/k_{B}T_{M}\right)},   
\end{align}   
where $\Lambda_{\mathrm{H}}$ is H $2s$-$1s$ two-photon decay rate, $\Lambda_{\mathrm{He}}$ is HeI $2^{1}s$-$1^{1}s$ two-photon decay rate, $\nu_{ps} \equiv \nu_{\mathrm{He}2^{1}p} - \nu_{\mathrm{He}2^{1}s}$, $K_{\mathrm{H}} = c^{3}/(8\pi \nu_{\mathrm{H}2p}^{3} H)$, and $K_{\mathrm{He}} = c^{3}/(8\pi \nu_{\mathrm{He}2^{1}p}^{3} H)$. $\alpha_{\mathrm{H}}$ and $\alpha_{\mathrm{He}}$ are the case B recombination coefficients and they are well fitted by
\begin{align}
\alpha_{\mathrm{H}} &= 10^{-19} F\frac{at^{b}}{1+ct^{d}} \left[ \mathrm{m^{3}s^{-1}} \right], \\
\alpha_{\mathrm{He}} &= q \left[\sqrt{\frac{T_{M}}{T_{2}}}\left(1+\frac{T_{M}}{T_{2}}\right)^{1-p} \left( 1+ \frac{T_{M}}{T_{1}}\right)^{1+p} \right]^{-1} \left[\mathrm{m^{3}s^{-1}}\right], 
\end{align}
with $t = T_{M}/10^{4} \left[ \mathrm{K}\right],\ a=4.309,\ b=-0.6166,\ c=0.6703,\ d=0.5300,\ F=1.14$ and $q = 10^{-16.744},\ p=0.711,\ T_{1} = 10^{5.114} \left[ \mathrm{K}\right],\ T_{2} = 3\left[ \mathrm{K}\right]$. Finally, $\beta_{\mathrm{H}}$ and $\beta_{\mathrm{He}}$ are the photoionization coefficients
\begin{align}
\beta_{\mathrm{H}} &= \alpha_{\mathrm{H}}\left(\frac{2\pi m_{e} k_{B}T_{M}}{h^{2}}\right)^{3/2}\exp \left(-\frac{h\nu_{\mathrm{H}2s}}{k_{B}T_{M}}\right), \\
\beta_{\mathrm{He}} &= \alpha_{\mathrm{He}}\left(\frac{2\pi m_{e} k_{B}T_{M}}{h^{2}}\right)^{3/2}\exp \left(-\frac{h\nu_{\mathrm{He}2^{1}s}}{k_{B}T_{M}}\right). 
\end{align}

Now, we show how the above quantities depend on $\alpha$. Since binding energies scale as $\alpha^{2}$, the frequencies $\nu$ also scale as $\alpha^{2}$, and $K_{\mathrm{H}} (K_{\mathrm{He}}) \propto \nu_{\mathrm{H}2p}^{-3}(\nu_{\mathrm{He}2^{1}p}^{-3})\propto \alpha^{-6}$. According to \cite{Kaplinghat:1998ry}, two-photon decay rates $\Lambda_{\mathrm{H}}$ and $\Lambda_{\mathrm{He}}$ scale as $\alpha^{8}$, and the recombination coefficients $\alpha_{\mathrm{H}}$ and $\alpha_{\mathrm{He}}$ scale as $\alpha^{2(1+\xi)}$, where we adopt $\xi=0.7$ following \cite{Kaplinghat:1998ry}.

\begin{figure}
\centering
\includegraphics[width=12cm]{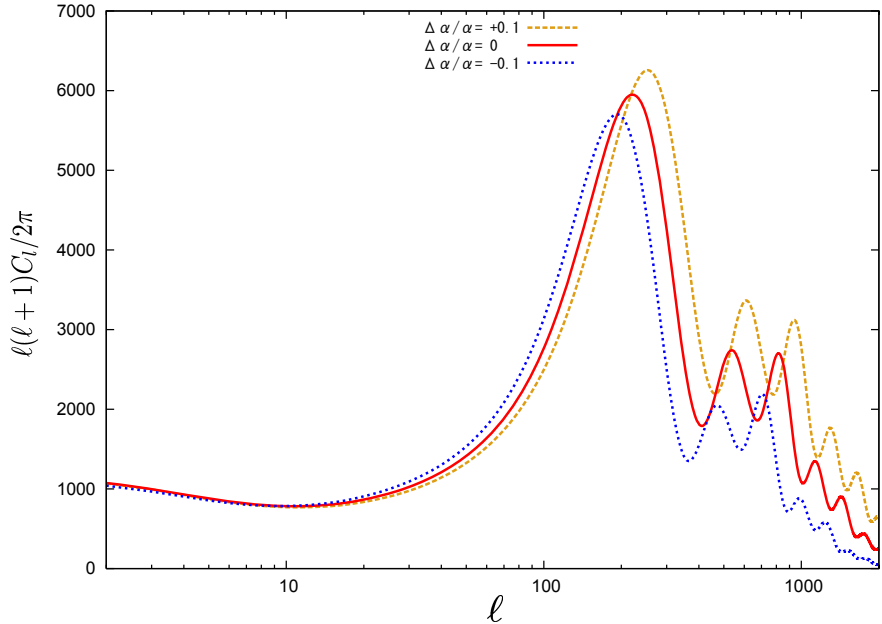}
\caption{The CMB temperature anisotropy spectra for no change of $\alpha$ (solid red curve), an increase of $\alpha$ by 10$\%$ (dashed yellow curve), a decrease of $\alpha$ by 10$\%$ (dotted blue curve). }
\label{fig:wmap}
\end{figure}

As already pointed out in \cite{Hannestad:1998xp,Kaplinghat:1998ry,Ichikawa:2006nm}, the larger value of $\alpha$ at the recombination epoch results in the higher redshift of the last scattering surface. Thus, increasing $\alpha$ result in three characteristic signatures in the angular power spectrum of the temperature anisotropy, namely, shift of the peaks to higher multipoles, increase of the height of the peaks due to the enhanced early integrated Sachs Wolfe effect, and decrease of the small-scale diffusion damping effect. These features can be seen in FIG.\ref{fig:wmap}.

 We constrain the variation of $\alpha$ using three kinds of CMB anisotropy spectra, namely, angular power spectrum of temperature anisotropy, $C_{\ell}^{TT}$, that of E-mode polarization, $C_{\ell}^{EE}$, and cross correlation of temperature and E-mode polarization $C_{l}^{TE}$ of the 5-year WMAP data \cite{Nolta:2008ih,Dunkley:2008ie,Komatsu:2008hk}. For this purpose, we have modified the CAMB code \cite{Lewis:1999bs,camb} including the RECFAST code to calculate the theoretical anisotropy specra for different values of $\alpha$ at recombination and we performed the parameter estimation using Markov-Chain Monte-Carlo (MCMC) techniques implemented in the CosmoMC code \cite{Lewis:2002ah,cosmomc}.   

 We have run the CosmoMC code on four Markov chains. To check the convergence, we used the "variance of chain means"/"mean of chain variances" $R$ statistic and adopted the condition $R-1 < 0.03$
 
 First, we consider the modified version of the flat $\Lambda$CDM model, that is, as for cosmological parameters we take $(\Omega_{B}h^{2},\Omega_{DM} h^{2},H_{0},n_{s},A_{s},\tau,\Delta\alpha/\alpha)$, where $\Omega_{B}h^{2}$ is the normalized baryon density, $\Omega_{DM}h^{2}$ is the normalized cold-dark-matter density, $H_{0} \equiv 100h \left[\mathrm{km\ sec^{-1} \ Mpc^{-1}} \right]$ is the Hubble constant, $n_{s}$ is the spectral index of the primordial curvature perturbation, $A_{s}$ is its amplitude, and $\Delta\alpha/\alpha \equiv (\alpha_{\mathrm{rec}} -\alpha_{\mathrm{now}})/\alpha_{\mathrm{now}}$ is the variation of the fine structure constant at recombination relative to its present value. We have also analized the standard flat $\Lambda$CDM model without $\Delta\alpha/\alpha$  and compared the other parameter values between these two models. 

\begin{table}
\caption{MCMC results on the mean values and 68$\%$ confidence intervals of cosmological parameters for the two models.}
\label{hyou1}
\begin{center}
\begin{tabular}{|l|c|c|}
\hline 
  & modified flat $\Lambda$CDM & standard flat $\Lambda$CDM \\
\hline \hline
$100\Omega_{B}h^{2}$ & $2.27^{+0.10}_{-0.10}$ & $2.27^{+0.06}_{-0.05}$ \\ \hline
$\Omega_{DM}h^{2}$ & $0.109^{+0.006}_{-0.006}$ & $0.109^{+0.006}_{-0.006}$ \\ \hline
$\tau$ & $0.0877^{+0.007}_{-0.008}$ & $0.0876^{+0.007}_{-0.008}$ \\ \hline
$n_{s}$ & $0.966^{+0.014}_{-0.015}$ & $0.964^{+0.014}_{-0.014}$ \\ \hline 
$\log(10^{10} A_{s})$ & $3.06^{+0.05}_{-0.04}$ & $3.06^{+0.04}_{-0.04}$ \\ \hline
$H_{0}$ & $71.9^{+7.0}_{-7.1}$ & $72.1^{+2.6}_{-2.6}$ \\ \hline
$\Delta\alpha/\alpha$ & $-0.000894^{+0.0131}_{-0.0148}$ & $-$ \\ 
\hline
\end{tabular}
\end{center}
\end{table}

\begin{figure}
\includegraphics[width=16cm]{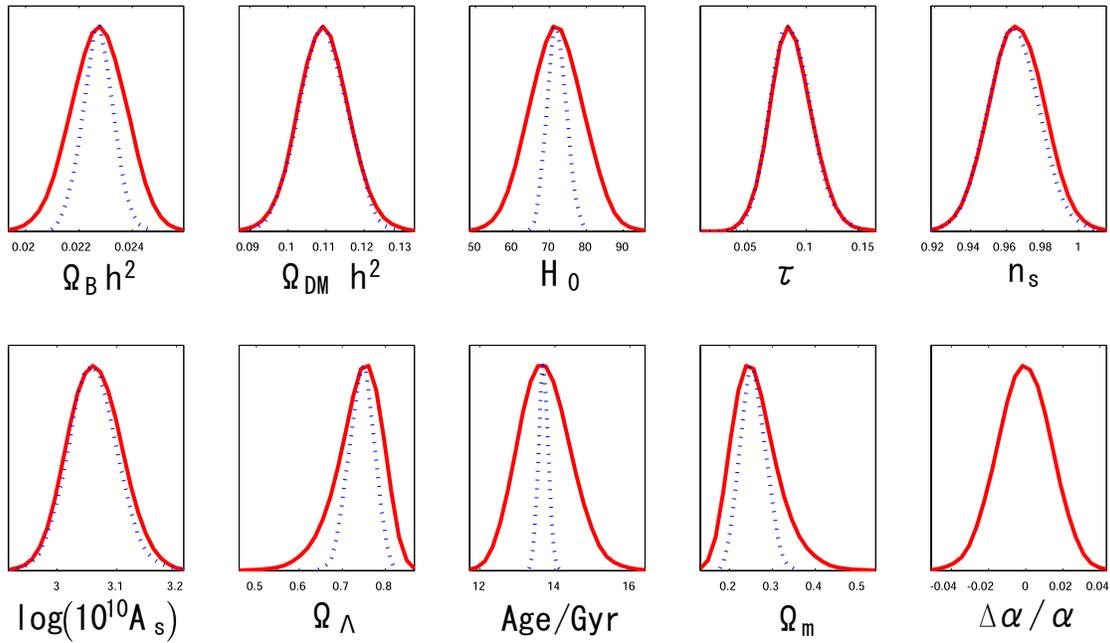}
\caption{One-dimensional marginalized posterior distributions for the parameters of the modified flat $\Lambda$CDM model (solid red curve), and for the standard flat $\Lambda$CDM model (dotted blue curve).}
\label{fig:1dim1}
\end{figure}

\begin{figure}
\centering
\includegraphics[width=16cm]{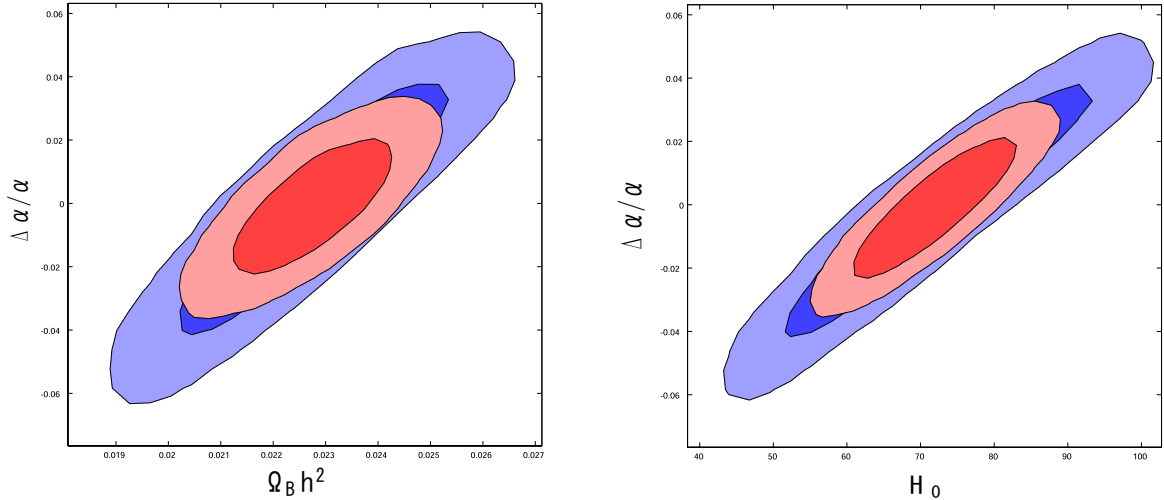}
\caption{Two-dimensional marginalized posterior distributions for $\Delta\alpha/\alpha$-$\Omega_{B}h^{2}$ and $\Delta\alpha/\alpha$-$H_{0}$. Red contours are with HST prior, and blue contours are without it. For each contour pairs, the inner one represent the $68\%$ bound and the outer one $95\%$.}
\label{fig:2dim1}
\end{figure}

\begin{figure}
\centering
\includegraphics[width=16cm]{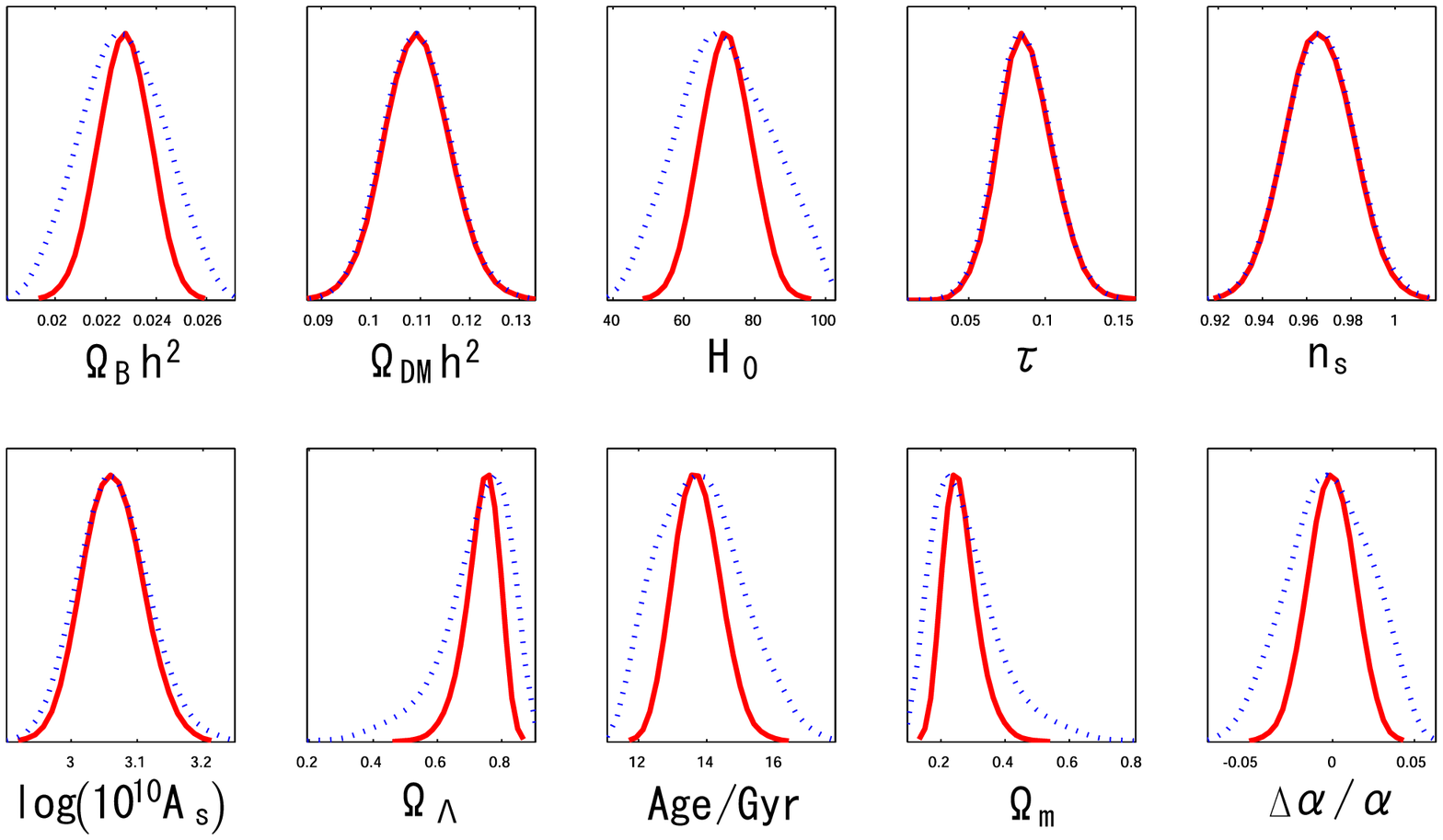}
\caption{One-dimensional marginalized posterior distributions for the results with HST prior (solid red curve), and without it (dotted blue curve). }
\label{fig:1dim2}
\end{figure}

 The results obtained from MCMC calculations are given in TABLE \ref{hyou1}, and FIG.\ref{fig:1dim1}. In TABLE \ref{hyou1}, we present the mean values and the 68$\%$ confidence intervals of the cosmological parameters and FIG. \ref{fig:1dim1} shows the one-dimensional marginalized posterior distributions of the parameters. From these results, it can be seen that the effect of the additional parameter $\Delta\alpha/\alpha$ is only to increase the errors of the other parameters and the mean values of the other parameters in modified flat $\Lambda$CDM are practically the same as in the standard flat $\Lambda$CDM. The marginalized distributions of $H_{0}$ and $\Omega_{B}$ in FIG.\ref{fig:1dim1} suggest the degeneracy of these parameters with $\Delta\alpha/\alpha$. 

Actually, in the above calcutions, we have incorporated the result of Hubble Key Project of the Hubble Space Telescope (HST) on the Hubble parameter $H_{0}$ \cite{Freedman:2000cf}, that is, we have imposed a prior that $H_{0}$ is a Gaussian with the mean $72\left[\mathrm{km\ sec^{-1} \ Mpc^{-1}} \right]$ and the variance $8\left[\mathrm{km\ sec^{-1} \ Mpc^{-1}} \right]$. If we do not use the HST prior, we can only get weaker constraints on the parameter values because of projection degeneracy. To check the effect of the HST prior, we show two-dimensional marginalized distributions with and without HST prior in FIG.\ref{fig:2dim1}. and one-dimensional distributions in FIG.\ref{fig:1dim2}. It is confirmed that the HST prior is very important in order to realistically constrain the time varition of $\alpha$.

The 95$\%$ confidence interval and the mean value of $\Delta\alpha/\alpha$ from 5-year WMAP data with HST prior are
\begin{equation}
-0.028 < \Delta\alpha/\alpha < 0.026 \ \ \mathrm{and} \ \ \Delta\alpha/\alpha = -0.000894,
\end{equation}
respectively. Without the HST prior, they read 
\begin{equation}
-0.050 < \Delta\alpha/\alpha < 0.042\ \ \mathrm{and}\ \ \Delta\alpha/\alpha = -0.00181,
\end{equation}
respectively. Previous results from 1-year WMAP data are $-0.048 < \Delta\alpha/\alpha < 0.032$ and $ -0.107<\Delta\alpha/\alpha<0.043$, with and without HST prior, respectively \cite{Ichikawa:2006nm}, so our results from 5-year WMAP data are about 30$\%$ tighter than those from 1-year WMAP data. We also note that for the 1-year data the mean value of $\Delta\alpha/\alpha$  was found to be $\Delta\alpha/\alpha=-0.04$ without the HST prior although it was practically equal to 0 with it. For the 5-year data we have found that the mean value remains practically intact whether we use the HST prior or not. This may be interpreted as an indication that the observational cosmology has made a step forward to the concordance at an even higher level.

\begin{figure}
\centering
\includegraphics[width=16cm]{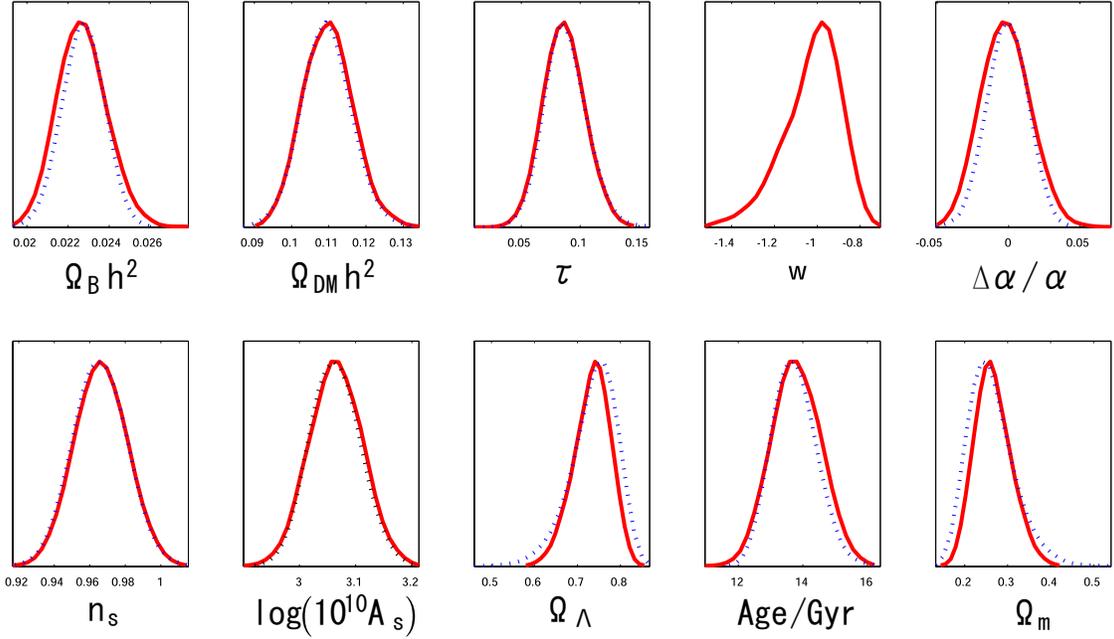}
\caption{One-dimensional marginalized posterior distributions for the parameters with variable $w$ (solid red curve), and $w=-1$ (dotted blue curve).}
\label{1dim3}
\end{figure}

 Next, we take $(\Omega_{B}h^{2},\Omega_{DM} h^{2},H_{0},n_{s},A_{s},\tau, w, \Delta\alpha/\alpha)$, where $w$ is the dark energy equation of state. In addition to the 5-year WMAP data, we use the HST and Supernova Legacy Survey \cite{Astier:2005qq} priors here. Although the possibility of the degeneracy of $w$ with $\Delta\alpha/\alpha$ was pointed out some time ago \cite{Huey:2001ku}, we can conclude that such a degeneracy is very weak now, for we have both temperature and polarization data enough to constrain $w$ and $\Delta\alpha/\alpha$ simultaneously (see FIG.\ref{1dim3}). The 95$\%$ confidence interval and the mean value of $\Delta\alpha/\alpha$ in this case are 
\begin{equation}
-0.033 < \Delta\alpha/\alpha < 0.032\ \ \mathrm{and}\ \ \Delta\alpha/\alpha= -0.00186
\end{equation}
which is slightly larger compared with the model with the cosmological constant.
 
In summary, in terms of the MCMC analysis using CosmoMC code, we have updated constraints on the time variation of the fine structure constant $\alpha$ based on 5-year WMAP data. We obtained tighter constraints compared with previous results from 1-year WMAP data due to the inclusion of the polarization data and the decrease of the statistical errors. Compared with the result based on the 3-year WMAP data \cite{Stefanescu:2007aa}, where comparison between the cases with and without HST prior has not been made, the resultant limit are almost of the same order of magnitude but the mean value of ours is closer to 0. We have verified that the null result is favored about the variation of $\alpha$ and the addition of this new parameter $\Delta\alpha/\alpha$ does not essentially affect the determinations of the other standard parameters contrary to the case of the analysis based on the 1-year WMAP data \cite{Ichikawa:2006nm}.       

We have also studied the possibility of the degeneracy between $w$ and $\Delta\alpha/\alpha$, finding the slightly relaxed limit on the latter parameter due to the addition of $w$, but no drastic degeneracy. Also in this case, we cannot find any evidence of time varying $\alpha$ in the 5-year WMAP CMB data.

\bigskip

\section*{Note added}
After we have finished our analysis, we bacame aware of a paper by Scoccola,
Landau, and Vucetich \cite{Scoccola:2008jw}, who also analyzed constraints on the time variation of $\alpha$ using 5-year WMAP data. Their main focus, however, is dependence on the details of the recombination scenario, which they have
shown to be weak. It appears that they did not incorporate HST prior in
contrast to our analysis.  Hence our paper is complementary to theirs.

\section*{Acknowledgements}
MN would like to thank Kiyotomo Ichiki and Takeshi Chiba for discussions. We would also like to thank Professor John D. Barrow for his useful comments on the first version of this manuscript. This work was partially supported by JSPS Grant-in-Aid for Scientific Research No.~19340054.

\end{document}